\documentclass[onecolumn]{article}
\input{epsf.sty}

\title{Depolarization of multiple scattered light in atmospheres due to anisotropy of small grains and molecules }

\author{N. A. Silant'ev\thanks{E-mail:nsilant@bk.ru}\,\,, G. A. Alekseeva,\, V. V. Novikov
\medskip\\ Central Astronomical Observatory at Pulkovo of Russian Academy of Sciences,\\ 196140,
Saint-Petersburg, Pulkovskoe shosse 65, Russia}

\begin{document}

\maketitle

\begin{abstract}
Freely oriented small anisotropic grains and molecules depolarize radiation both in single scattering and
in the process of multiple scattering. Especially large depolarization occurs for resonant scattering
corresponding to the electron transitions between the energy levels with very different quantum numbers.
The existence of light absorption also changes essentially the angular distribution and polarization of radiation,
outgoing from an atmosphere. In the present paper we consider these effects in detail both for continuum radiation and
for resonant lines. Because the term describing the depolarization deals with isotropic radiation, we consider
the axially symmetric part of radiation.
We derived the formulas for observed intensity and polarization
using the invariance-principles both for continuum and resonant scattering. We confine ourselves to two problems -
the diffuse reflection of the light beam from semi-infinite atmosphere, and the Milne problem.

\end{abstract}

\section{Introduction}

 The observation of polarization of  radiation emanating from atmospheres of planets, stars,  stellar envelopes
and accretion discs  gives additional information {\bf on} these objects. First of all, the polarization
demonstrates the {\bf existence of} various types of anisotropy in observing objects. The observed polarization helps us
to construct various models of the objects, {\bf namely} the models of non-spherical atmospheres. If we observe the
eclipsing binary, the value of observed polarization is {\bf a} variable. In all {\bf these} cases we have to know the local distribution
of polarization of emanating radiation in an atmosphere. The calculation of polarization emanating from
semi-infinite plane-parallel atmosphere is one of the basic problem in the theory of radiative transfer.

In most papers
the scattering particles (molecules, dust grains) are assumed to be small compared to the wavelength of radiation.
In this case the dipole scattering is most important. The incident radiation induces the time dependent dipole
moment, which is the source of scattered light. The induced moment depends on the structure of grain or molecule.
If the scattering particle is anisotropic, the induced dipole moment depends on the orientation of particle. Only
for isotropic particle (say, electron) the induced moment is always the same. Scattering on isotropic particle
gives rise to the maximal polarization of radiation. The ensemble of chaotically oriented anisotropic grains or
molecules gives rise to smaller polarization as compared to the case of isotropic particles. Thus, anisotropic
structure of grains or molecules depolarizes scattered radiation. In the present paper, we restrict ourselves
to dipole approximation.

Clearly, the radiative transfer in an atmosphere
having the anisotropic grains or molecules is more difficult for consideration than {\bf the} usually assumed scattering on
isotropic particles. On the other hand, the real atmospheres consist of anisotropic particles. For this reason, the
estimation of depolarizing effect in {\bf such} atmospheres is interesting and important. Below we present the solution of
standard problems of radiative transfer (see Chandrasekhar 1960) taking into account the effect of depolarization.
We consider the diffuse reflection of light beam from semi-infinite atmosphere without sources of radiation
, and the Milne problem, where the sources of radiation are located far from the surface ($\tau \gg 1$).
The solutions of these problems, according to Chandrasekhar (1960), follow from the invariance-principle and
the radiative transfer equation without source term.

The axially symmetric part of radiation is described by two intensities - $I_l(\tau,\mu,\nu)$ and $I_r(\tau,\mu,\nu)$.
Here $\tau $ is the optical depth below the surface of semi-infinite plane-parallel atmosphere,
$\mu =\cos\vartheta$ with $\vartheta$ being the angle between the outer normal ${\bf N}$ to the surface and the direction
of light propagation ${\bf n}$, $\nu$ is the frequency of light.
 The intensity $I_l$ describes the light linearly polarized in the plane (${\bf nN}$), and $I_r$ has polarization
perpendicular to this plane. The total intensity $I=I_l+I_r$, and the Stokes parameter $Q=I_l-I_r$. The Stokes
parameter $U\equiv 0$. The degree of linear polarization is equal to $p=|I_l-I_r|/(I_l+I_r)$.
Circularly polarized light is described by {\bf a} separate equation. We do not consider this equation.

Introducing the (column) vector ${\bf I}$ with the components ($I_l, I_r$) we obtain the following matrix
transfer equation for multiple scattering of continuum radiation on small anisotropic particles
(see Chandrasekhar 1960; Dolginov et al. 1995):
\[
\mu \frac{d{\bf I}(\tau,\mu)}{d\tau}= {\bf I}(\tau,\mu)-
\frac{1-q}{2}\int\limits_{-1}^{1}d\mu'\,\hat P(\mu,\mu'){\bf I}(\tau,\mu'),
\]
\begin{equation}
\hat P(\mu,\mu')=\left (\begin{array}{rr} P_1, P_2\\ P_3, P_4\end{array}\right ) =
\overline{b}_1\frac{3}{4}\left(\begin{array}{rr}2(1-\mu^2)(1-\mu'^2)+\mu^2\mu'^2\,\,\,\,\,, \,\,\,\,\,\mu^2\\
 \,\,\,\,\,\mu'^2\,\,\,\,\,\,\,,\,\,\,\,\,\,\, 1\end{array}\right)
 +\overline{b}_2\frac{3}{2}\left (\begin{array}{rr}1 , 1 \\ 1, 1\end{array}\right).
\label{1}
\end{equation}

\noindent Here $q=N_a\sigma_a/(N_s\sigma_s+N_a\sigma_a)$ is the probability of light absorption, $N_a$ and $N_s$ are
the number densities of absorbing (the grains) and scattering particles, respectively;
$d\tau=(N_s\sigma_s+N_a\sigma_a)dz$ determines the dimensionless optical depth, $\sigma_s$ and $\sigma_a$ are the
cross sections of scattering and absorption.

The scattering cross section of small particles (dust grains, molecules) is
$\sigma_s=(8\pi/3)(\omega/c)^4(b_1+3b_2)$, where $\omega=2\pi\nu$ is cyclic frequency of light, $c$ is the speed of
light. For freely (chaotic{\bf ally}) oriented particles $\sigma_s$ is independent of the polarization of incident
electromagnetic
wave ${\bf E}(\omega)$. The values $b_1$ and $b_2$ are related to polarizability tensor $\beta_{ij}(\omega)$ of
a particle as a whole. Induced dipole moment of a particle, as whole, is equal to
$p_i(\omega)=\beta_{ij}(\omega)E_j(\omega)$. Anisotropic particle with axial symmetry is characterized by two
polarizabilities - along the symmetry axis $\beta_{\parallel}(\omega)$, and in transverse direction
$\beta_{\perp}(\omega)$. For such particles

\[
b_1=\frac{1}{9}|2\beta_{\perp}+\beta_{\parallel}|^2+\frac{1}{45}|\beta_{\parallel}-\beta_{\perp}|^2,
\]
\begin{equation}
b_2=\frac{1}{15}|\beta_{\parallel}-\beta_{\perp}|^2.
\label{2}
\end{equation}
\noindent In transfer equation we use the dimensionless parameters
\begin{equation}
\overline{b}_1=\frac{b_1}{b_1+3b_2},\,\,\, \overline{b}_2=\frac{b_2}{b_1+3b_2},\,\,\,\overline{b}_1+3\overline{b}_2=1.
\label{3}
\end{equation}
\noindent For needle like particles ($|\beta_{\parallel}|\gg |\beta_{\perp}|)$ parameters $\overline{b}_1=0.4,
\overline{b}_2=0.2$, and for plate like particles ($|\beta_{\perp}|\gg |\beta_{\parallel}|$) we have
$\overline{b}_1=0.7, \overline{b}_2=0.1$.

Parameter $\overline{b}_2$ describes the depolarization of radiation, scattered by freely oriented anisotropic
particles. The integral term in Eq.(1) (the source function ${\bf B}(\tau,\mu)=(B_l(\tau,\mu), B_r(\tau,\mu))$ for
single scattering of non-polarized  radiation ($I_{l,r}=(1/2)I_0\delta(\mu-\mu_0)$) acquires the form:

\[
B_l(\mu)=\frac{3}{16}(1-q)I_0\{[2(1-\mu^2)(1-\mu_0^2)+\mu^2(1+\mu_0^2)]\overline{b}_1+4\overline{b}_2\},
\]
\begin{equation}
B_r(\mu){\bf =}\frac{3}{16}(1-q)I_0[(1+\mu_0^2 )\overline{b}_1+4\overline{b}_2].
\label{4}
\end{equation}
\noindent The degree of polarization of single scattered radiation is then given by

\begin{equation}
p(\mu,\mu_0)=\frac{B_l(\mu)-B_r(\mu)}{B_l(\mu)+B_r(\mu)}=\frac{\overline{b}_1(1-\mu^2)(1-3\mu_0^2)}
{[2(1-\mu^2)(1-\mu_0^2)+(1+\mu^2)(1+\mu_0^2)]\overline{b}_1+8\overline{b}_2}.
\label{5}
\end{equation}
\noindent The peak polarization is reached for $\mu=0$, i.e., for scattering of radiation parallel
to the surface of an atmosphere:

\begin{equation}
p_{max}=\frac{\overline{b}_1(1-3\mu_0^2)}{(3-\mu_0^2)\overline{b}_1+8\overline{b}_2}.
\label{6}
\end{equation}
\noindent  When $\mu_0=1$ the scattering angle is equal to $90^{\circ}$ and the polarization degree (6) takes
the value $p_{max}=-\overline{b}_1/(\overline{b}_1+4\overline{b}_2)$. For needle like particles $p_{max}=-33.3\%$,
and for plate like particles $p_{max}=- 63.63\%$. For isotropic particles $(\overline{b}_2=0)$ one has
$p_{max}=-100\%$. The minus sign denotes that preferable oscillations of wave electric vector are perpendicular
to meridional plane $({\bf nN})$.

Note that in Chandrasekhar (1960) the parameter $\gamma$ is used. Our parameters $\overline{b}_1$ and
$\overline{b}_2$ are related to $\gamma $ according to formulas: $\overline{b}_1=(1-\gamma)/(1+2\gamma)$ and
$\overline{b}_2=\gamma/(1+2\gamma)$.

Usually the light polarization weakly influences the angular distribution of radiation, emanating from an atmosphere.
Thus, in conservative Milne's problem the ratio $J(\mu)=I(0,\mu)/I(0,0)$ takes the value 3.06, if the polarization terms
in transfer equation are taken into account. If these terms are omitted the value $J(\mu)=3.02$. For this reason,
if one is interested in the intensity of radiation, then one uses the scalar transfer equation only for
intensity $I(\tau,\mu)$. This equation in our axially symmetric case can be derived from matrix equation (1)
through the summation ($I_l+I_r$) and substituting $I_l(\tau,\mu')=I_r(\tau,\mu')=I(\tau,\mu')/2$ in the integrand of
Eq.(1). As a result, we obtain the standard equation:

\[
\mu\frac{dI(\tau\mu)}{d\tau}=I(\tau,\mu)-\frac{1-q}{2}\int\limits_{-1}^{1}d\mu'\,P(\mu,\mu')I(\tau,\mu'),
\]

\begin{equation}
P(\mu,\mu')=\left[\frac{1}{8}(3-\mu^2)(3-\mu'^2)+\mu^2\mu'^2\right]\overline{b}_1+3\overline{b}_2.
\label{7}
\end{equation}
\noindent The integrand $\hat P(\mu,\mu')$ in Eq.(7) is the sum of products of type $\varphi(\mu)\varphi(\mu')$. According to
Chandrasekhar (1960), it is possible to derive a system of non-linear integral equations for three H-functions.
These functions can be used in the derivation of the formulas for outgoing intensity of reflected radiation,
and in the Milne problem.

The phase matrix $\hat P(\mu,\mu')$ can also be presented analogous to scalar phase function
(see Lenoble 1970; Abhyankar \& Fymat 1971). In our case ($\overline{b}_2\neq 0$) the phase matrix
 $\hat P(\mu,\mu')$ can be rewritten in the form:

\begin{equation}
\hat P(\mu,\mu')\equiv \hat P(\mu^2,\mu'^2)=\overline{b}_1 \hat M(\mu^2)\hat M(\mu'^2)^T
+\frac{3}{4}\overline{b}_2\hat L \hat {L}^T.
\label{8}
\end{equation}

\begin{equation}
\hat M(\mu^2)=\frac{\sqrt{3}}{2}
\left (\begin{array}{ll}\mu^2\,\, ,\, (1-\mu^2)\sqrt{2}\\ \,1\,\,\,\,\,,\,0\,\end{array}\right ),\,\,\,
\hat L=\left (\begin{array}{cc}\, 1\,,\,1\,\\ \,1\,,\,1\,\end{array}\right ).
\label{9}
\end{equation}

\noindent Here the superscript T stands for matrix transpose.

In the present paper, we consider in detail two basic problems of the radiative transfer theory - the reflection of polarized
light from semi-infinite plane parallel atmosphere, and the Milne problem corresponding to the thermal sources in
very deep layers of an absorbing atmosphere.

The main features of our investigation are consideration of problems with depolarization parameter
$\overline{b}_2$, and
taking into account the true absorbtion (parameter $q$) (i.e., the existence of absorbing grains in an atmosphere.)
Firstly we consider the solution of transfer equation only for intensity $I$, and then the more complex case of
matrix equation for intensities $I_l$ and $I_r$. Recall that the depolarization parameter
$\overline{b}_2$ gives the contribution to the axially symmetric part of radiation.

A detailed consideration of problems without the parameter $\overline{b}_2$ is presented in many papers (see,
for example, Chandrasekhar 1960; Horak \& Chandrasehar 1961; Lenoble 1970; Abhyankar \& Fymat 1971).

Below we present briefly the standard general description of the problems under consideration and
then turn to particular solutions.

\subsection{The case of resonant scattering}

For investigation of multiple scattering of resonant radiation one uses the matrix transfer equation of
the general form (see, for example, Hummer 1962; Ivanov et al. 1997a, 1997b; Dementyev 2008):

\begin{equation}
\mu\frac{d{\bf I}(\tau,\mu,\nu)}{d\tau}=\alpha(\nu){\bf I}(\tau,\mu,\nu)-\frac{1-q}{2}\int\limits_{-1}^{1}
d\mu'\int\limits_{-\infty}^{\infty}d\nu'\,\hat P(\mu,\nu;\mu',\nu'){\bf I}(\tau,\mu',\nu'),
\label{10}
\end{equation}
\noindent where $\nu$ is the frequency of light, the absorbtion factor in a resonant line is $\alpha_{resonant}(\nu)=
\alpha_0\,\varphi(\nu)$, the optical depth $d\tau =\alpha_0dz$ takes into account the mean absorbtion factor in
a line, the dimensionless factor $\alpha(\nu)=\varphi(\nu)+
\alpha_{cont}/\alpha_0$, with $\alpha_{cont}$ being the extinction factor in nearby continuum.

The normalized function $\varphi(\nu)$ describes the form of line. Often one uses the limiting
forms - Gaussian or Doppler

\begin{equation}
\varphi(\nu)=\frac{1}{\sqrt{\pi}\Delta\nu_D}\exp{\left[-\left ( \frac{\nu-\nu_0}{\Delta\nu_D}\right )^2\right]},
\label{11}
\end{equation}
\noindent and the Lorentz

\begin{equation}
\varphi(\nu)=\frac{\delta}{\pi}\,\cdot \frac{1}{(\nu-\nu_0)^2+\delta^2}.
\label{12}
\end{equation}
\noindent The function $\varphi(\nu)$ is normalized to unity, namely:

\[
\int\limits_{-\infty}^{\infty}d\nu\,\varphi(\nu)=1.
\]

 Here $\nu_0$ is central frequency of a resonant line, $\Delta\nu_D$ is Doppler width of a line:

\[
\Delta\nu_D^2=\Delta\nu_{th}^2+\Delta\nu_{turb}^2=\frac{\nu_0^2}{c^2}(u_{th}^2+u_{turb}^2),
\]
\noindent where the thermal velocity is determined by the temperature $u_{th}^2=2k_BT/m$, and the turbulent velocity
is determined as a mean value of chaotic macroscopic motions $u_{turb}^2=\langle u^2({\bf r},t)\rangle$. The
value $\delta $ depends on  widths of the atomic energy levels.
The Gaussian form of a line arises as a result of Doppler frequency shifts due to thermal and turbulent motions of
atoms and molecules. This form usually corresponds to the {\bf line core}. The Lorentz form characterize the
wings of a line which are often blanketed by line environment.

The matrix $\hat P(\mu,\nu;\mu',\nu')$ in general has very complex form (see McKenna 1985; Landi Degl'Innocenti \&
Landolfi 2004). This is the reason why one uses model of fully redistributed frequencies:

\begin{equation}
\hat P(\mu,\nu;\mu',\nu')=\varphi(\nu)\varphi(\nu')\hat P(\mu,\mu'),
\label{13}
\end{equation}

\noindent where the matrix $\hat P(\mu,\mu')$ has the form (8). This means that the scattering law in {\bf a}
spectral line formally coincides with that by scattering on anisotropic freely oriented small particles. The parameters
$\overline{b}_1$ and $\overline{b}_2$ in this case are related to the parameters
 $E_1$ and $E_2$ used in Chandrasekhar (1960) as $\overline{b}_1=E_1$ and $\overline{b}_2=E_2/3$
with $E_1+E_2=1$.
 It should be noted that in this case the depolarization parameter
$\overline{b}_2$ can be very large and play{\bf s} a very important role in the calculation of resonant emission
polarization.
Large depolarization of spectral lines is both the consequence of chaotic orientations and the mixture of the
dipole electron transitions with very different quantum numbers.
Thus, the estimates demonstrate that for $H\alpha $ line, consisting of 5 close components, the value
$E_2\simeq 0.3$. Note, that the Doppler broadening of close components overlap the frequency differences between
them and the line, as a whole, can be considered as a single line having the Doppler form (Varshalovich et
al. 2006; Lekht et al. 2008).

From the matrix equation (10) one can also derive the separate scalar transfer equation for intensity
$I(\tau,\mu,\nu)$:

\begin{equation}
\mu\frac{dI(\tau,\mu,\nu)}{d\tau}=\alpha(\nu)I(\tau,\mu,\nu)-\frac{1-q}{2}\int\limits_{-1}^{1}d\mu'\,\int\limits_{-\infty}^
{\infty}d\nu'\, P(\mu,\nu;\mu'\nu')I(\tau,\mu',\nu'),
\label{14}
\end{equation}
\noindent where the scalar function $P(\mu,\nu;\mu'\nu')=(P_1+P_2+P_3+P_4)/2$ is equal to the sum
of all four components of matrix $\hat P(\mu,\nu;\mu',\nu')$.

\section{Basic formulas}

In this section we briefly recall the basic theoretical formulas first derived by Chandrasekhar (1960). This is done
as a matter of convenience in consideration of particular problems depending on depolarization parameter
$\overline{b}_2$ and the degree of true absorption $q$.

\subsection{The equation for scattering matrix}
Let a parallel beam of light with fluxes $\pi F_l$ and $\pi F_r$ along the direction characterized
by $\cos\vartheta_0=-\mu_0$ and azimuth angle $\varphi_0$ be incident on
 the surface of a semi-infinite plane-parallel atmosphere.
The intensity of the light diffusely reflected from the atmosphere can be expressed in terms of scattering
matrix:

\begin{equation}
{\bf I}(0,\mu,\varphi)=\frac{1}{4\mu}\hat S(\mu,\varphi;\mu_0,\varphi_0){\bf F}.
\label{15}
\end{equation}

The scattering matrix $\hat S(\mu,\varphi ;\mu_0,\varphi_0)$ obeys the matrix equation (the invariance principle):

\[
\left (\frac{1}{\mu}+\frac{1}{\mu_0}\right )\hat S(\mu,\varphi;\mu_0,\varphi_0)=(1-q)\left [\hat P(\mu,\varphi;-\mu_0,
\varphi_0)+\right.
\]

\[
\left.\frac{1}{4\pi}\int\limits_0^1\frac{d\mu''}{\mu''}\,\int\limits_0^{2\pi}d\varphi''\,\hat P(\mu,\varphi;\mu'',
\varphi'')\hat S(\mu'',\varphi'';\mu_0,\varphi_0)+\frac{1}{4\pi}\int\limits_0^1\,\frac{d\mu'}{\mu'}\int\limits_0^{2\pi}
d\varphi'\,\hat S(\mu,\varphi ;\mu',\varphi')\hat P(-\mu',\varphi';-\mu_0,\varphi_0)+\right.
\]

\begin{equation}
\left.\frac{1}{(4\pi)^2}
\int\limits_0^1\frac{d\mu'}{\mu'}\,\int\limits_0^{2\pi}d\varphi'\int\limits_0^1\frac{d\mu''}{\mu''}\,\int\limits
_0^{2\pi}d\varphi''\hat S(\mu,\varphi ;\mu',\varphi')\,\hat P(-\mu',\varphi'; \mu'',\varphi'')\, \hat S(\mu'',\varphi'';
\mu_0,\varphi_0)\right ].
\label{16}
\end{equation}

In our axially symmetric case the dependence on azimuthal angles will be absent. The integrations over $\varphi'$ and
$\varphi''$ give the values $2\pi$, and formula (16) takes a simpler form. In formula (16) we take the common
factor $(1-q)$ out of brackets, i.e. $\hat S\sim (1-q)$. Note, that in our case $\hat P(\mu,\mu')\equiv \hat
P(\mu^2,\mu'^2)$.

If the matrix $\hat P(\mu,\mu')$ has the form

\begin{equation}
\gamma_1\hat M_1(\mu^2)\hat M_1(\mu'^2)^T +\gamma_2 \hat M_2(\mu^2)\hat M_2(\mu'^2)^T,
\label{17}
\end{equation}
\noindent then the expression for $\hat S(\mu,\mu')$ takes the following form:

\begin{equation}
\hat S(\mu,\mu') =\frac{(1-q)\mu\mu'}{\mu+\mu'}\left [ \gamma_1\hat N_1(\mu)\hat N_1(\mu')^T  +
\gamma_2\hat N_2(\mu)\hat N_2(\mu')^T \right],
\label{18}
\end{equation}
\noindent where $\hat N_n(\mu)$ obeys the relation:

\begin{equation}
\hat N_n(\mu)=\hat M_n(\mu^2)+\frac{1}{2}\int\limits_0^1\,\frac{d\mu'}{\mu'}\hat S(\mu,\mu')\hat M_n(\mu'^2).
\label{19}
\end{equation}
\noindent Substitution of Eq.(18) into Eq.(19) gives rise to the system of non-linear matrix equations
for matrices $\hat N_1$ and $\hat N_2$. It should be noted that the expression (18) is valid only when
$\hat M_n(\mu )=\hat M_n(\mu^2)$. In general case the expression for $\hat S(\mu,\mu')$ takes
the form:

\begin{equation}
\hat S(\mu,\mu')=\frac{(1-q)\mu\mu'}{\mu+\mu'}[\,\gamma_1\hat R_1(\mu)\hat K_1(\mu')+
\gamma_2\hat R_2(\mu)\hat K_2(\mu')].
\label{20}
\end{equation}
\noindent The matrices $\hat R_n(\mu)$ and $\hat K_n(\mu)$ obey the equations:

\[
\hat R_n(\mu)=\hat M_n(\mu)+\frac{1}{2}\int\limits_0^1\frac{d\mu'}{\mu'}\hat S(\mu,\mu')\hat M_n(-\mu'),
\]

\begin{equation}
\hat K_n(\mu)=\hat M_n^T(-\mu)+\frac{1}{2}\int\limits_0^1\frac{d\mu'}{\mu'}\hat M_n^T(\mu')\hat S(\mu',\mu).
\label{21}
\end{equation}

The system of non-linear matrix equations for $\hat R_n$ and $\hat K_n$ are more complex than those for
$\hat N_n$.

\subsection{Formulas for Milne's problem}

The Milne problem deals with the solution of Eq.(1) when the sources of thermal radiation are placed in deep
layers of an atmosphere. The important part of this problem is the solution of the transfer equation in infinite
atmosphere.
This solution has the form ${\bf I}(\tau,\mu)={\bf g}(\mu)\exp{(k\tau )}/(1-k\mu)$. The (column) vector ${\bf g}(\mu)$
obeys the homogeneous matrix equation:

\begin{equation}
{\bf g}(\mu) = \frac{1-q}{2}\int\limits_{-1}^1 d\mu'\,\hat P(\mu,\mu')\frac{{\bf g}(\mu')}{1-k\mu'}.
\label{22}
\end{equation}

For phase matrix $\hat P(\mu^2,\mu'^2)$ and phase function $P(\mu^2,\mu'^2)$ the (column) {\bf vector} ${\bf g(\mu)}$
and the analogous scalar $g(\mu)$ depend on $\mu^2$. {\bf In  this case Eq.(22)  acquires more simple form:}

\[
{\bf g}(\mu^2)=(1-q)\int\limits_0^1\,d\mu'\,\hat P(\mu^2,\mu'^2)\frac{{\bf g}(\mu'^2)}{1-k^2\mu'^2}.
\]
\noindent Below we consider only such cases.

The homogeneous equation (22) has the solution, if the constant parameter $k$ obeys the characteristic equation
(zero's determinant of Eq.(22)). The most simple form of the characteristic equation is obtained for
the scalar transfer equation for intensity {\bf $I(\tau,\mu)=g(\mu)\exp{(k\tau)}/(1-k\mu)$} with the isotropic phase
function ($\overline{b}_1=0, \overline{b}_2=1/3$):

\begin{equation}
(1-q)f_0(k)=(1-q)\frac{1}{2k}\ln{\frac{1+k}{1-k}}=1.
\label{23}
\end{equation}
\noindent For the case of Rayleigh  phase function ($\overline{b}_1=1,\overline{b}_2=0$) the characteristic
equation is:

\begin{equation}
1-\frac{3}{8}(1-q)(3f_0+3f_4-2f_2)+\frac{9}{8}(1-q)^2(f_0f_4-f_2^2)=0.
\label{24}
\end{equation}
\noindent Here the functions $f_n(k)$ are determined as:

\[
f_0(k)=\int\limits_0^1\frac{d\mu}{1-k^2\mu^2}=\frac{1}{2k}\ln{\frac{1+k}{1-k}},
\]
\[
f_2(k)=\int\limits_0^1d\mu \,\frac{\mu^2}{1-k^2\mu^2}=\frac{f_0-1}{k^2},
\]
\begin{equation}
f_4(k)=\int\limits_0^1d\mu \,\frac{\mu^4}{1-k^2\mu^2}=\frac{3f_2-1}{3k^2}.
\label{25}
\end{equation}
\noindent For Rayleigh scattering ($\overline{b}_1=1,\overline{b}_2=0$) with taking into account the
polarization terms the characteristic equation is:

\begin{equation}
1-\frac{3}{4}(1-q)(3f_0+3f_4-4f_2)+\frac{9}{8}(1-q)^2(f_0^2-f_2^2+2f_0f_4-2f_0f_2)=0.
\label{26}
\end{equation}

In Tables 1 and 2 we present the values $k(q)$ for these three cases. From these tables, we see that
the $k$ - values for isotropic
scattering are larger than those corresponding to Rayleigh  phase function. The inclusion of polarization
terms gives rise to smaller values of  $k$ compared to case where they are neglected. The maximum relative difference
of $\simeq 1\%$ between the $k$-values obtained from Eq.(23) and Eq.(24) occurs at $q\simeq 0.4$, while that
between Eq.(24) and Eq.(25) occurs at $q\simeq 0.2$. For small absorption ($q\ll 1$) the value of $k\simeq \sqrt{3q}$.
This approximation for $k$ is valid up to $q\simeq 0.05$, where the relative difference with the exact value
is $\simeq 2\%$.

The invariance-principles give rise to the formula (Chandrasekhar, 1960):

\begin{equation}
{\bf I}(0,\mu)=Const \left [\frac{{\bf g}(\mu)}{1-k\mu}-\frac{1}{2\mu}\int\limits_0^1\frac{d\mu'}{1+k\mu'}\,
\hat S(\mu,\mu'){\bf g}(\mu')\right ].
\label{27}
\end{equation}
\noindent The value {\bf $Const$} is related with the total flux of outgoing radiation.

\section{The intensity $I(\tau,\mu)$ for an atmosphere with depolarization parameter $\overline{b}_2$}

The transfer equation for $I(\tau,\mu)$ is presented in Eq.(7). The phase function has the form:

\begin{equation}
P(\mu,\mu')=\frac{1}{8}\overline{b}_1 (3-\mu^2)(3-\mu'^2)+\overline{b}_1\mu^2\,\mu'^2 +3\overline{b}_2.
\label{28}
\end{equation}
\noindent For the case of unpolarized radiation, the general formulas (18) and (19), give the following
expression for $S(\mu,\mu')$:

\begin{equation}
S(\mu,\mu')=\frac{(1-q)\mu\mu'}{\mu+\mu'}\left [\frac{1}{8}\overline{b}_1\psi(\mu)\psi(\mu')+\overline{b}_1\phi
(\mu)\phi(\mu')+3\overline{b}_2\xi(\mu)\xi(\mu')\right ],
\label{29}
\end{equation}
\noindent where the functions $\psi(\mu),\phi(\mu)$ and $\xi(\mu)$ are expressed in term $S(\mu,\mu')$:

\[
\psi(\mu)=(3-\mu^2)+\frac{1}{2}\int\limits_0^1 \frac{d\mu'}{\mu'}\,(3-\mu'^2)\,S(\mu,\mu'),
\]
\[
\phi(\mu)=\mu^2 +\frac{1}{2}\int\limits_0^1 \frac{d\mu'}{\mu'}\,\mu'^2\,S(\mu,\mu'),
\]

\begin{equation}
\xi(\mu)=1+\frac{1}{2}\int\limits_0^1 \frac{d\mu'}{\mu'}\,S(\mu,\mu').
\label{30}
\end{equation}

From the above equations, we see that $\xi(\mu)=[\psi(\mu)+\phi(\mu)]/3$. For convenience in the following
operations, we introduce new notations:

\begin{equation}
\psi(\mu)=4[A(\mu)+B(\mu)],\,\, \phi(\mu)=2A(\mu)-B(\mu),\,\,\xi(\mu)=2A(\mu)+B(\mu).
\label{31}
\end{equation}

The system of equations for $A(\mu)$ and $B(\mu)$ is the following:

\[
A(\mu)=\frac{1+\mu^2}{4}+\frac{3}{8}(1-q)\,\mu \int\limits_0^1 d\mu'\,\frac{1+\mu'^2}{\mu+\mu'}T(\mu,\mu'),
\]

\begin{equation}
B(\mu)=\frac{1-\mu^2}{2}+\frac{3}{4}(1-q)\,\mu \int\limits_0^1 d\mu'\,\frac{1-\mu'^2}{\mu+\mu'}T(\mu,\mu'),
\label{32}
\end{equation}
\noindent  where

\begin{equation}
T(\mu,\mu')=\overline{b}_1\,(2AA'+BB')+\overline{b}_2\,(2A+B)(2A'+B')\equiv T(\mu',\mu).
\label{33}
\end{equation}
\noindent Here and in what follows, for brevity, we use the notations $A(\mu)=A, A(\mu')=A'$ etc.
In new notations the scattering function $S(\mu,\mu')$ acquires the form:

\begin{equation}
S(\mu,\mu')=\frac{3(1-q)\,\mu\,\mu'}{\mu+\mu'}\,T(\mu,\mu').
\label{34}
\end{equation}

The functions $A(\mu)$ and $B(\mu)$ terms of one $H$-function. We now briefly describe the way
to achieve this. From the system of equations (32) we obtain:

\begin{equation}
2A(\mu)+B(\mu)=1+\frac{3}{2}(1-q)\,\mu \int\limits_0^1 d\mu'\frac{T(\mu,\mu')}{\mu+\mu'}.
\label{35}
\end{equation}
\noindent  Similarly we find the expression for $2A(\mu)-B(\mu)$ and using formula (35), one can obtain
the relation:

\begin{equation}
2A(\mu)\phi_A(\mu)=B(\mu)\phi_B(\mu),
\label{36}
\end{equation}
\noindent where the functions $\phi_A(\mu)$ and $\phi_B(\mu)$ are:

\[
\phi_A(\mu)=1-\mu^2+\frac{3}{2}(1-q)\mu\,\{\overline{b}_1(\mu A_0-A_1)+\overline{b}_2\,[\,\mu (2A_0+B_0)-(2A_1+B_1)]\},
\]

\begin{equation}
\phi_B(\mu)=1+\mu^2-\frac{3}{2}(1-q)\mu\,\{\overline{b}_1(\mu B_0-B_1)+\overline{b}_2\,[\,\mu (2A_0+B_0)-(2A_1+B_1)]\}.
\label{37}
\end{equation}
\noindent Here we use the notations

\begin{equation}
A_n=\int\limits_0^1 d\mu\,\mu^n\,A(\mu),\,\,\,\, B_n=\int\limits_0^1 d\mu\,\mu^n\,B(\mu).
\label{38}
\end{equation}

The expression (36) is valid if there exist the relations:

\begin{equation}
A(\mu)=\frac{1}{4}\phi_B(\mu)\,H(\mu),\,\,\,\, B(\mu)=\frac{1}{2}\phi_A(\mu)\,H(\mu).
\label{39}
\end{equation}

Introducing these relations in Eq.(35), one can obtain the following equation for H - function:

\[
\frac{1}{2}(\phi_A+\phi_B)\,H=1+\frac{3}{16}(1-q)\,\mu\,\int\limits_0^1\,\frac{d\mu'}{\mu+\mu'}\,[\,\overline{b}_1
(2\phi_A\phi'_A+\phi_B\phi'_B)+
\]

\begin{equation}
2\overline{b}_2(\phi_A+\phi_B)(\phi'_A+\phi'_B)]\,H\,H'.
\label{40}
\end{equation}

The functions $\phi_A(\mu)$ and $\phi_B(\mu)$ are polynomials of the type $\phi=a+b\mu+c\mu^2$. Hence one
can show that

\begin{equation}
\frac{\phi(\mu)}{\mu+\mu'}=\frac{\phi(-\mu')}{\mu+\mu'}+b+c(\mu-\mu').
\label{41}
\end{equation}

Substituting Eq.(41) for $\phi_A(\mu)$ and $\phi_B(\mu)$ in Eq.(40), and keeping in mind that
terms with moments $A_0, A_1, B_0$, $B_1$  and $\mu^2$  in the left side of Eq.(40) are canceled by the same terms
in the right side of this equation, one can obtain the standard formula for H - function:

\begin{equation}
H(\mu)=1+\mu\,H(\mu) \int\limits_0^1 d\mu'\frac{\Psi(\mu')}{\mu+\mu'}H(\mu'),
\label{42}
\end{equation}
\noindent where

\[
\Psi(\mu)=\frac{3}{16}(1-q)\overline{b}_1[2\phi_A(\mu)\phi_A(-\mu)+\phi_B(\mu)\phi_B(-\mu)]+
\]
\begin{equation}
\frac{3}{8}(1-q)\overline{b}_2\,[\phi_A(\mu)+\phi_B(\mu)][\phi_A(-\mu)+\phi_B(-\mu)].
\label{43}
\end{equation}

For deriving Eq.(42) we have used the relations:

\[
2A_0+B_0=1+\frac{3}{4}(1-q)\,[\,\overline{b}_1\,(2A_0^2+B_0^2)+\overline{b}_2\,(2A_0+B_0)^2\,],
\]

\begin{equation}
2A_0-B_0=\frac{1}{3}+\frac{3}{4}(1-q)\,[\,\overline{b}_1\,(2A_1^2+B_1^2)+\overline{b}_2\,(2A_1+B_1)^2\,],
\label{44}
\end{equation}
\noindent which follows from Eq.(32), if one uses the symmetry relation $T(\mu,\mu')=T(\mu',\mu)$ and the
substitution $\mu \to \mu'$ and the opposite $\mu'\to \mu$. The initial double integral over $\mu$ and $\mu'$ coincides with
that after substitutions.  As a result, in the nominator of the sum of these integrals
arises the term $(\mu+\mu')$ which cancels the same term in the denominator.

The explicit form of the function $\Psi(\mu)$ is the following:

\[
\Psi(\mu)=\frac{3}{16}(1-q)\{\,3\overline{b}_1+8\overline{b}_2 - \overline{b}_1\,[\,2-(1-q)\overline{b}_1-
3(1-q)\overline{b}_2\,]\,\mu^2+
\]
\begin{equation}
 3\overline{b}_1\,[\,1-(1-q)\overline{b}_1-3(1-q)\overline{b}_2\,]\,\mu^4 \}.\,\,\,\,
\label{45}
\end{equation}
\noindent This formula follows from Eq.(43) after the substitution of expressions (37) for $\phi_A(\mu)$ and
$\phi_B(\mu)$ and using the relations (44). It is of interest that we do not use the explicit values
of $A_n$ and $B_n$. It seems such situation reflects some symmetry properties of the transfer equation (7).

For $\overline{b}_1=0$ and $\overline{b}_2=1/3$ the expression (45) gives $\Psi(\mu)=(1-q)/2$. In this case
Eq.(42) coincides with the standard H -function equation for isotropic scattering. For $\overline{b}_2=0$ and
$\overline{b}_1=1$ the $\Psi(\mu)$ - function is given by

\begin{equation}
\Psi(\mu)=\frac{3}{16}(1-q)[\,3-(1+q)\mu^2 +3q\mu^4\,].
\label{46}
\end{equation}
\noindent For conservative atmosphere $(q=0)$ the expression (46) transforms into the known form, namely
$\Psi(\mu)=(3/16)(3-\mu^2)$.

Taking into account the relations (39), one can write the scattering function $S(\mu,\mu')$ in the form:

\[
S(\mu,\mu')=\frac{3(1-q)\mu\,\mu'}{8(\mu+\mu')}H(\mu)H(\mu')\times
\]
\begin{equation}
[\,\overline{b}_1(2\phi_A\phi'_A+\phi_B\phi'_B)+
2\overline{b}_2(\phi_A+\phi_B)(\phi'_A+\phi'_B)\,].
\label{47}
\end{equation}
\noindent The expression in brackets is polynomial of $\mu$ and $\mu'$.

\subsection{The Milne problem }

The function $g(\mu)$ for Milne problem with depolarization parameter $\overline b_2$ takes the form

\begin{equation}
g(\mu)=\frac{3}{16}(1-q)\,\int\limits_{-1}^1\frac{d\mu'}{1-k\mu'}[\,\overline{b}_1(3-\mu^2-\mu'^2+3\mu^2\,\mu'^2)+
8\overline{b}_2\,]\,g(\mu').
\label{48}
\end{equation}
\noindent According to this equation

\begin{equation}
g(\mu)=g_0+g_2\,\mu^2.
\label{49}
\end{equation}

The homogeneous system of equations for parameters $g_0$ and $g_2$ has the form:

\[
\left \{\,1-\frac{3}{8}(1-q)[(3f_0-f_2)\overline{b}_1+8\overline{b}_2f_0]\right \}g_0-\frac{3}{8}(1-q)
[(3f_2-f_4)\overline{b}_1+8\overline{b}_2f_2\,]\,g_2=0,
\]
\begin{equation}
-\frac{3}{8}(1-q)\overline{b}_1(3f_2-f_0)\,g_0+[\,1-\frac{3}{8}(1-q)\overline{b}_1(3f_4-f_2)]\,g_2=0.
\label{50}
\end{equation}
\noindent The $f_n$ - functions are defined in Eq.(25). The characteristic equation (the zero value for determinant
of system (50)) is:

\begin{equation}
1-\frac{3}{8}(1-q)[\,\overline{b}_1\,(3f_0+3f_4-2f_2)+8\overline{b}_2\,f_0]+
\frac{9}{8}(1-q)^2\overline{b}_1\,(\overline{b}_1+3\overline{b}_2)\,(f_0f_4-f_2^2)=0.
\label{51}
\end{equation}
\noindent The solution of this algebraic equation determines the parameter $k$.

The formula (27) for the intensity of outgoing radiation takes the form:

\begin{equation}
I(0,\mu)=\frac{Const}{1-k\mu}\left [g(\mu)-\frac{3}{2}(1-q) \int\limits_0^1 d\mu'\frac{\mu'(1-k\mu)}
{(\mu+\mu')(1+k\mu')}T(\mu,\mu')\,g(\mu')\,\right ].
\label{52}
\end{equation}
\noindent Taking into account the equality

\begin{equation}
\frac{\mu'(1-k\mu)}{(\mu+\mu')(1+k\mu')}=\frac{1}{1+k\mu'}-\frac{\mu}{\mu+\mu'}
\label{53}
\end{equation}
\noindent and Eq.(35) for $(2A+B)$, the expression (52) can be presented as follows:

\[
I(0,\mu)=\frac{Const}{1-k\mu}\left \{\,[2A(\mu)+B(\mu)]\,g(\mu)+\frac{3}{2}(1-q)\,\mu\,
[\,T_1(\mu)-\mu T_0(\mu)\,]\,g_2\,- \right.
\]

\begin{equation}
\left. \frac{3}{2}(1-q)\int\limits _0^1\,d\mu'\frac{T(\mu,\mu')}{1+k\mu'}\,g(\mu')\right \},
\label{54}
\end{equation}
\noindent  where we have introduced the notations:

\begin{equation}
T_n(\mu)=\int\limits_0^1\,d\mu'\,\mu'^n\,T(\mu,\mu').
\label{55}
\end{equation}

If one uses the relations of $A(\mu)$ and $B(\mu)$  with $H$ - function (see Eq.(39)), then the main
term $I(0,\mu)\sim H(\mu)/(1-k\mu)$ appears.

\section{{\bf The intensities $I_l$ and $I_r$ for an atmosphere with depolarization parameter} $\overline{b}_2$}

The system of equations for $I_l(\tau,\mu)$ and $I_r(\tau,\mu)$  is described by the matrix transfer equation (1).
According to Eqs.(8) and (9), the matrix $\hat P(\mu,\mu')$ can be presented as the sum of product
of matrices $\hat M(\mu^2)$
and $\hat M^T(\mu'^2)$, and the product of $\hat L$ and $\hat L^T=\hat L$. The general theory (see, Eqs.(17) - (21))
gives the following form of scattering matrix $\hat S(\mu,\mu')$:

\begin{equation}
\hat S(\mu,\mu')\equiv \left ( \begin{array}{ll} S_1\,,\, S_2\\ S_3\,,\, S_4 \end{array}\right )=
\frac{(1-q)\mu\,\mu'}{\mu+\mu'}
\left [\,\overline{b}_1\hat N(\mu)\hat N^T(\mu')+
\frac{3}{4}\,\overline{b}_2\,\hat \phi(\mu)\hat \phi^T(\mu')\right ],
\label{56}
\end{equation}
\noindent where the matrices $\hat N(\mu)$ and $\hat \phi(\mu)$ are related to $\hat S(\mu,\mu')$ as:

\begin{equation}
\hat N(\mu)=\left (\begin{array}{ll}\,N_1\, , \,N_2\,\\\,N_3\,, \,N_4\,\end{array}\right )=
\hat M(\mu)+\frac{1}{2}\int\limits_0^1\frac{d\mu'}{\mu'}\hat S(\mu,\mu')\hat M(\mu'^2),
\label{57}
\end{equation}

\begin{equation}
\hat \phi(\mu)\equiv \left( \begin{array}{ll}\,a\,,\,a\,\\ \,b\,,\,b\,\end{array}\right )=
\hat L+\frac{1}{2}\int\limits_0^1\frac{d\mu'}{\mu'}\hat S(\mu,\mu')\hat L.
\label{58}
\end{equation}
\noindent The properties $\phi_1=\phi_2=a$ and $\phi_3=\phi_4=b$ are consequences of explicit form of
matrix $\hat L$ ($L_{ik}=1$, see Eq.(9)). Eq.(58) can be written as

\[
a(\mu)=1+\frac{1}{2}\int\limits_0^1\,\frac{d\mu'}{\mu'}[\,S_1(\mu,\mu')+S_2(\mu,\mu')\,],
\]

\begin{equation}
b(\mu)=1+\frac{1}{2}\int\limits_0^1\,\frac{d\mu'}{\mu'}[\,S_3(\mu,\mu')+S_4(\mu,\mu')\,].
\label{59}
\end{equation}
\noindent It is easy to check that

\begin{equation}
a=2\left (\frac{N_1}{\sqrt{3}}+\frac{N_2}{\sqrt{6}}\right ),\,\,
b=2\left (\frac{N_3}{\sqrt{3}}+\frac{N_4}{\sqrt{6}}\right ).
\label{60}
\end{equation}
\noindent So, the matrix $\hat \phi(\mu)$ is expressed in terms of matrix $\hat N(\mu)$.

Let us introduce more convenient notations:

\[
A=\frac{N_1+N_3}{2\sqrt{3}},\,\,\,\,B=\frac{N_2+N_4}{\sqrt{6}},
\]

\begin{equation}
C=\frac{N_3-N_1}{2\sqrt{3}},\,\,\,\,D=\frac{N_4-N_2}{\sqrt{6}}.
\label{61}
\end{equation}
\noindent In these notations the scattering matrix {\bf $\hat S(\mu,\mu')$} acquires the form:

\[
\hat S(\mu,\mu')= \frac{3}{2}\frac{(1-q)\mu\,\mu'}{\mu+\mu'}\times
\]
\[
\left \{[\,\overline{b}_1\,(2AA'+BB')+\overline{b}_2\,(2A+B)(2A'+B')] \left (\begin{array}{ll}\,1\,,\,1\,\\
\,1,\,1\,\end{array}\right) + \right.
\]
\[
\left. [\,\overline{b}_1\,(2CC'+DD')+\overline{b}_2\,(2C+D)(2C'+D')] \left (\begin{array}{ll}\,\,1\,,-1\,\\
-1\,,\,\,1\,\end{array}\right)+ \right.
\]
\[
\left. [\,\overline{b}_1\,(2AC'+BD')+\overline{b}_2\,(2A+B)(2C'+D')] \left (\begin{array}{ll}\,-1\,,\,1\,\\
\,-1\,,\,1\,\end{array}\right)+ \right.
\]
\begin{equation}
\left.[\,\overline{b}_1\,(2CA'+DB')+\overline{b}_2\,(2C+D)(2A'+B')] \left (\begin{array}{ll}\,-1\,,\,-1\,\\
\,\,\,\,\,1\,,\,\,\,\,\,1\,\end{array}\right)\right \}.
\label{62}
\end{equation}

The transition to the case of scalar equation, considered in Section 3, can be made according to formula
$S(\mu,\mu')=(S_1+S_2+S_3+S_4)/2$. As a result, we recover to Eqs.(34) and (33), where the functions $A(\mu)$
and $B(\mu)$ obey the system of equations (32).

The substitution of expression (56) into Eqs.(57) and (58) gives rise to explicit form of system
 of non-linear equations for
functions $A(\mu),B(\mu),C(\mu)$ and $D(\mu)$:

\[
A(\mu)=\frac{1+\mu^2}{4}+\frac{3}{8}(1-q)\mu\,\int\limits_0^1\frac{d\mu'}{\mu+\mu'}[\,(1+\mu'^2)T(\mu,\mu') +
(1-\mu'^2)R(\mu,\mu')\,],
\]
\[
B(\mu)=\frac{1-\mu^2}{2}+\frac{3}{4}(1-q)\mu\,\int\limits_0^1\frac{d\mu'}{\mu+\mu'}[\,(1-\mu'^2)T(\mu,\mu')-
(1-\mu'^2)R(\mu,\mu')\,],
\]
\[
C(\mu)=\frac{1-\mu^2}{4}+\frac{3}{8}(1-q)\mu\,\int\limits_0^1\frac{d\mu'}{\mu+\mu'}[\,(1+\mu'^2)R(\mu',\mu)+
(1-\mu'^2)U(\mu,\mu')\,],
\]
\begin{equation}
D(\mu)=-\frac{1-\mu^2}{2}+\frac{3}{4}(1-q)\mu\,\int\limits_0^1\frac{d\mu'}{\mu+\mu'}[\,(1-\mu'^2)R(\mu',\mu)-
(1-\mu'^2)U(\mu,\mu')\,].
\label{63}
\end{equation}
\noindent Here we used the notations:
\[
T(\mu,\mu')=\overline{b}_1\,(2AA'+BB')+\overline{b}_2\,(2A+B)(2A'+B')=T(\mu',\mu),
\]
\[
R(\mu,\mu')=\overline{b}_1\,(2AC'+BD')+\overline{b}_2\,(2A+B)(2C'+D'),
\]
\begin{equation}
U(\mu,\mu')=\overline{b}_1\,(2CC'+DD')+\overline{b}_2\,(2C+D)(2C'+D')=U(\mu',\mu).
\label{64}
\end{equation}
\noindent If we neglect the polarization terms ($C(\mu)=0, D(\mu)=0$), then the system (63) transforms to
the system (32) for functions $A$ and $B$.

For unpolarized incident radiation $(I_l(-\mu')=I_r(-\mu')=F_0/2)$ the degree of polarization at $\tau=0$
is (see Eq.(15)):

\begin{equation}
p(\mu,\mu')=\frac{I_l-I_r}{I_l+I_r}=-\frac{\overline{b}_1(2CA'+DB')+\overline{b}_2\,(2C+D)(2A'+B')}{T(\mu,\mu')}.
\label{65}
\end{equation}
\noindent Recall that, for brevity, we use the notations $A=A(\mu), A'=A(\mu')$ etc. For the same reason
in formula (65) we take $\mu'=\mu_0$, where $\mu_0$ characterizes the incident radiation at the surface
(see section 2). Substitution in Eq.(65)
instead of functions $A,B,C$ and $D$  the corresponding free terms in equations (63), gives rise to the expression
(5).

It is of interest that the function{\bf s} $A, B, C$ and $D$ depend on one function, which we denote as
$H_0(\mu)$. Let us briefly prove this
statement. From the system (63) one can obtain the expression:

\begin{equation}
2A+B=1+\frac{3}{2}(1-q)\mu\int\limits_0^1\frac{d\mu'}{\mu+\mu'}T(\mu,\mu').
\label{66}
\end{equation}
\noindent Deriving the difference $2A-B$ and using Eq.(66), we obtain the relation between $A$ and $B$:

\begin{equation}
2A(\mu)\phi_A(\mu)=B(\mu)\phi_B(\mu),
\label{67}
\end{equation}
\noindent where

\[
\phi_A(\mu)=1-\mu^2 -\frac{3}{2}(1-q)\mu\,[\,\overline{b}_1\,(A_1-\mu\,A_0+c_0-c_2)+\overline{b}_2\,\beta(\mu)],
\]
\[
\phi_B(\mu)=1+\mu^2+\frac{3}{2}(1-q)\mu\,[\, \overline{b}_1\,(B_1-\mu B_0+d_0-d_2)+\overline{b}_2\,\beta(\mu)],
\]
\begin{equation}
\beta(\mu)=2(A_1-\mu A_0)+(B_1-\mu B_0)+2(c_0-c_2)+d_0-d_2.
\label{68}
\end{equation}
\noindent Here we introduce the notations:

\begin{equation}
c_n(\mu)=\int\limits_0^1\,\frac{d\mu'}{\mu+\mu'}\,C(\mu')\mu'^n.
\label{69}
\end{equation}
\noindent Analogous formula is defined for $a_n(\mu), b_n(\mu)$ and $d_n(\mu)$. The relation (67) is
 valid if

\begin{equation}
A(\mu)=\frac{1}{4}\phi_B(\mu)H_0(\mu),\,\,\,\, B(\mu)=\frac{1}{2}\phi_A(\mu)H_0(\mu).
\label{70}
\end{equation}

Deriving the sum $2C+D$, we obtain the relation analogous to Eq.(67):

\begin{equation}
2C(\mu)\phi_C(\mu)=D(\mu)\phi_D(\mu),
\label{71}
\end{equation}
\noindent where

\[
\phi_C(\mu)=1-\frac{3}{2}(1-q)\mu\,[\,\overline{b}_1\,a_0+\overline{b}_2\,(2a_0+b_0)],
\]
\begin{equation}
\phi_D(\mu)=-1+\frac{3}{2}(1-q)\mu\,[\,\overline{b}_1\,b_0+\overline{b}_2\,(2a_0+b_0)].
\label{72}
\end{equation}
\noindent The relation (71) implies that

\begin{equation}
C(\mu)=\frac{1}{4}\phi_D(\mu)H_1(\mu),\,\,\,\,D(\mu)=\frac{1}{2}\phi_C(\mu)H_1(\mu).
\label{73}
\end{equation}

Eq.(66) can also be written as follows:

\begin{equation}
2A(\mu)\phi_C(\mu)=1 +B(\mu)\phi_D(\mu).
\label{74}
\end{equation}
\noindent Solving the system of equations (71) and (74), one can obtain $\phi_C(\mu)$ and $\phi_D(\mu)$ as
functions of $A, B, C$ and $D$:

\[
\phi_C=\frac{D}{2(AD-BC)}\equiv\frac{2D}{H_1},
\]
\begin{equation}
\phi_D=\frac{2C}{2(AD-BC)}\equiv\frac{4C}{H_1}.
\label{75}
\end{equation}
\noindent It follows from Eqs.(75) that

\begin{equation}
H_1(\mu)=4[\,A(\mu)D(\mu)-B(\mu)C(\mu)\,].
\label{76}
\end{equation}

Derivation of $2C-D$ shows that

\begin{equation}
2C(\phi_A+\mu^2\phi_C)=1-\mu^2 +D(\phi_B+\mu^2\phi_D).
\label{77}
\end{equation}
\noindent Joint with Eq.(71) this formula gives rise to relation:

\begin{equation}
2C\phi_A=1-\mu^2+D\phi_B.
\label{78}
\end{equation}
\noindent Substitution of the equalities $\phi_A=2B/H_0$ and $\phi_B=4A/H_0$ in the above relation, and taking
into account the expression (76), gives the relation between $H_1(\mu)$ and $H_0(\mu)$:

\begin{equation}
H_1(\mu)=-(1-\mu^2)H_0(\mu).
\label{79}
\end{equation}

Thus, all the functions - $A, B, C$ and $D$ are expressed in terms of one function $H_0(\mu)$.
Recall, that in the case of scalar transfer equation (7) the substitution of the relations $A=\phi_B H/4$
and $B=\phi_A H/2$ into
Eq.(35) (this is analog of Eq.(66)) gives rise to standard nonlinear equation (42) for $H$ - function.
In the present case of polarized transfer, the functions $\phi_A$ and $\phi_B$ depend on functions
$c_0, c_2, d_0$ and $d_2$. In other words the functions
$\phi_A$ and $\phi_B$ for polarized case are not the polynomials as they were in scalar case (see
 the expressions (37)).
For this reason we could not obtain the closed equation for $H_0(\mu)$. Equation (66) can serve as
the basis for iteration method to calculate the $H_0$ - function. The detail consideration of this problem will be given
in a  future publication.

\subsection{ The Milne problem}

The matrix integral equation for the (column) vector ${\bf g}(\mu)=(g_l, g_r)$ is presented in Eq.(22), where
the phase matrix $\hat P(\mu,\mu')$ is given in Eqs.(8) and (9). It follows from this equation that

\begin{equation}
{\bf g}(\mu)={\bf g}_0+\mu^2{\bf g}_2 = \left (\begin{array}{c}a\\b\end{array}\right )
 +\mu^2\left (\begin{array}{c}c\\0\end{array}\right )\,\,\,\,.
\label{80}
\end{equation}
\noindent Homogeneous system of algebraic equations for values $a,b$ and $c$ have the form:

\[
[1-\alpha_1(2f_0-2f_2)-2\alpha_2\,f_0]\,a-2\alpha_2f_0\,b-2[\alpha_1(f_2-f_4)+\alpha_2\,f_2\,]\,c=0,
\]
\[
-(\alpha_1f_2+2\alpha_2\,f_0)\,a+(1-\alpha_1\,f_0-2\alpha_2\,f_0)\,b-(\alpha_1\,f_4+2\alpha_2\,f_2)\,c=0,
\]
\begin{equation}
-\alpha_1(3f_2-2f_0)\,a-\alpha_1\,f_0\,b+[1-\alpha_1(3\,f_4-2f_2\,)]\,c=0.
\label{81}
\end{equation}
\noindent Here, for brevity, we use the notations:

\begin{equation}
\alpha_1=\frac{3}{4}(1-q)\,\overline{b}_1,\,\,\,\, \alpha_2=\frac{3}{4}(1-q_)\,\overline{b}_2.
\label{82}
\end{equation}

The values of $f_n$ are given in Eq.(25). The characteristic equation (the zero value for determinant
$\Delta(k)$ of system (81)) allows us to calculate the value of parameter $k$.

\[
\Delta(k)=\Delta_1(k)-4\alpha_2\,f_0+6\alpha_1\,\alpha_2(f_0^2-f_2^2+2f_0f_4-2f_0f_2)=0,
\]
\begin{equation}
\Delta_1(k)=1-\alpha_1(3f_0+3f_4-4f_2)+2\alpha_1^2(f_0^2-f_2^2+2f_0f_4-2f_0f_2).
\label{83}
\end{equation}

For dipole scattering ($\overline{b}_2=0, \overline{b}_1=1$) the expression (83) transforms to Eq.(26).
For isotropic scattering ($\overline{b}_1=0, \overline{b}_2=1/3$) Eq.(83) reduces to the equation
$4\alpha_2\,f_0=1$,
which determines the parameter $k(q)$ for scalar transfer equation with isotropic phase function.

The angular distribution and the polarization of outgoing radiation ${\bf I}(0,\mu)$ is described by Eq.(27).
Taking into account the equality (53), this equation can be written as

\begin{equation}
{\bf I}(0,\mu)=\frac{Const}{1-k\mu}\left [{\bf g}(\mu)+\frac{1}{2}\int\limits_0^1\,\frac{d\mu'}{\mu'}\hat S(\mu,\mu')
{\bf g}(\mu)-\frac{1}{2\mu}\int\limits_0^1\,\frac{d\mu'}{\mu'}\cdot\frac{\mu+\mu'}{1+k\mu'}\,\hat S(\mu,\mu'){\bf g}(\mu')
\right ].
\label{84}
\end{equation}

\section{Polarization of resonant radiation in a model of full frequency redistribution}

In this model the transfer equation for ${\bf I}=(I_l,I_r)$ has the form:

\begin{equation}
\mu\frac{d{\bf I}(\tau,\mu,\nu)}{d\tau}=\alpha(\nu){\bf I}(\tau,\mu,\nu)-\frac{1-q}{2}\int\limits_{-1}^{1}
d\mu'\,\int\limits_{-\infty}^{\infty}d\nu'\,\varphi(\nu)\varphi(\nu')\hat P(\mu^2,\mu'^2){\bf I}(\tau,\mu',\nu').
\label{85}
\end{equation}
\noindent Recall that the absorbtion factor of resonant radiation is $\alpha_{resonant}(\nu)=\alpha_0\varphi(\nu)$,
the quantity $\alpha(\nu)=\varphi(\nu)+\alpha_{cont}/\alpha_0$, and $d\tau=\alpha_0\,dz$.
 Eq.(85) has been investigated in detail in several earlier papers (see e.g. Ivanov et al. 1997a, 1997b;
Dementyev 2008). For a more complete list of references we refer the reader to Faurobert-Scholl \& Frisch (1989)
 and reviews by Nagendra (2003) and Nagendra \& Sampoorna (2009).
As opposed to these authors, our consideration is based on the invariance-principles (see Eq.(16)) which are valid
both for the continuum radiation and for resonant one.
For resonant radiation the phase matrix in Eq.(16) is replaced by $\varphi(\nu)\varphi(\nu')\hat P(\mu^2,\mu'^2)$
 and in the integral terms there now appears the integration over frequencies.

In our axially symmetric case Eq.(15) transforms to

\begin{equation}
{\bf I}(0,\mu,\nu)=\frac{1}{4\mu}\hat S(\mu,\nu;\mu_0,\nu_i){\bf F}_0(\mu_0,\nu_i).
\label{86}
\end{equation}
\noindent Here $\nu_i$ is the arbitrary frequency in an incident flux ${\bf F_0}(\mu_0,\nu_i)$ of resonant radiation.
According to the invariance principle, the matrix $\hat S(\mu,\nu;\mu_0,\nu_i)$ obeys the equation:

\[
\left (\frac{\alpha(\nu)}{\mu}+\frac{\alpha(\nu_i)}{\mu_0}\right )\hat S(\mu,\nu;\mu_0,\nu_i)=(1-q)\left [
\,\varphi(\nu)\varphi(\nu_i)\hat P(\mu^2,\mu_0^2)+\right.
\]

\[
\left.\frac{1}{2}\int\limits_0^1\frac{d\mu''}{\mu''}\,\int\limits_{-\infty}^{\infty}\,d\nu''\,
\varphi(\nu)\varphi(\nu'')
\hat P(\mu^2,\mu''^2)\hat S(\mu'',\nu'';\mu_0,\nu_i)+\frac{1}{2}\int\limits_0^1\,\frac{d\mu'}{\mu'}\int\limits_
{-\infty}^{\infty}\,d\nu'\,\hat S(\mu,\nu ;\mu',\nu')\hat P(\mu'^2,\mu_0^2)\varphi(\nu')\varphi(\nu_i)+\right.
\]

\begin{equation}
\left.\frac{1}{4}
\int\limits_0^1\frac{d\mu'}{\mu'}\,\int\limits_{-\infty}^{\infty}\,d\nu'\,\int\limits_0^1\frac{d\mu''}{\mu''}\,
\int\limits
_{-\infty}^{\infty}\,d\nu''\,\hat S(\mu,\nu ;\mu',\nu')\,\hat P(\mu'^2, \mu''^2)\, \hat S(\mu'',\nu'';
\mu_0,\nu_i)\varphi(\nu')\varphi(\nu'')\right ].
\label{87}
\end{equation}
\noindent It is convenient to take out from the matrix $\hat S(\mu,\nu;\mu_0,\nu_i)$ the product
 $\varphi(\nu)\varphi(\nu_i)$:

\begin{equation}
\hat S(\mu,\nu;\mu_0,\nu_i)=\varphi(\nu)\varphi(\nu_i)\hat S_1(\mu,\nu;\mu_0,\nu_i).
\label{88}
\end{equation}
\noindent The matrix $\hat S_1(\mu,\nu;\mu_0,\nu_i)$ obeys the equation:

\[
\left (\frac{\alpha(\nu)}{\mu}+\frac{\alpha(\nu_i)}{\mu_0}\right )\hat S_1(\mu,\nu;\mu_0,\nu_i)=(1-q)\left [
\,\hat P(\mu^2,\mu_0^2)+\right.
\]

\[
\left.\frac{1}{2}\int\limits_0^1\frac{d\mu''}{\mu''}\,\int\limits_{-\infty}^{\infty}\,d\nu''\,\varphi^2(\nu'')
\hat P(\mu^2,\mu''^2)\hat S_1(\mu'',\nu'';\mu_0,\nu_i)+\frac{1}{2}\int\limits_0^1\,\frac{d\mu'}{\mu'}\int\limits_
{-\infty}^{\infty}\,d\nu'\,\hat S_1(\mu,\nu ;\mu',\nu')\hat P(\mu'^
2,\mu_0^2)\varphi^2(\nu')+\right.
\]

\begin{equation}
\left.\frac{1}{4}
\int\limits_0^1\frac{d\mu'}{\mu'}\,\int\limits_{-\infty}^{\infty}\,d\nu'\,\int\limits_0^1\frac{d\mu''}{\mu''}\,
\int\limits
_{-\infty}^{\infty}\,d\nu''\,\hat S_1(\mu,\nu ;\mu',\nu')\,\hat P(\mu'^2, \mu''^2)\, \hat S_1(\mu'',\nu'';
\mu_0,\nu_i)\varphi^2(\nu')\varphi^2(\nu'')\right ].
\label{89}
\end{equation}

According to general theory (see Eqs.(17) - (21)) one can derive the following formula
(here and in what follows, for brevity, we take $\mu_0=\mu'$ and $\nu_i=\nu'$):

\[
\left(\frac{\alpha(\nu)}{\mu}+\frac{\alpha(\nu')}{\mu'}\right )\hat S_1(\mu,\nu;\mu'\nu')=
\]
\begin{equation}
\left [\,\overline{b}_1\,\hat N(\mu,\nu)\hat N^T(\mu',\nu')
+\frac{3}{4}\overline{b}_2\,\eta(\mu,\nu)\eta^T(\mu',\nu')\right ],
\label{90}
\end{equation}
\noindent where the matrices $\hat N(\mu,\nu)$ and $\hat \eta(\mu,\nu)$ are related to
$\hat S_1(\mu,\nu;\mu',\nu')$ as:

\[
\hat N(\mu,\nu)=\hat M(\mu^2)+\frac{1}{2}\int\limits _0^1 \frac{d\mu'}{\mu'}\int\limits_{-\infty}^{\infty}
d\nu'\,\hat S_1(\mu,\nu;\mu',\nu')\varphi^2(\nu')\hat M(\mu'^2),
\]
\begin{equation}
\hat \eta(\mu,\nu)=\hat L+\frac{1}{2}\int\limits_0^1\frac{d\mu'}{\mu'}\,\int\limits_{-\infty}^{\infty}d\nu' \hat
S_1(\mu,\nu;\mu',\nu')\varphi^2(\nu')\hat L.
\label{91}
\end{equation}
\noindent The formulas (91) are analogous to Eqs.(57) and (58) with additional factor $\varphi^2(\nu')$ and the
integration over $\nu'$. The matrix $\hat \eta$ has two independent components $\eta_1=\eta_2=a$ and $\eta_3=\eta_4=b$.
The functions $a$ and $b$ are related with components of $\hat N(\mu,\nu)$ according to Eq.(60). Using our
convenient {\bf notations} (61), we obtain the scattering matrix $\hat S$ in the form:

\begin{equation}
\hat S(\mu,\nu;\mu',\nu')=\frac{3}{2}\cdot\frac{(1-q)\mu\,\mu'}{\alpha(\nu')\mu+\alpha(\nu)\mu'}
\varphi(\nu)\varphi(\nu')\{\}.
\label{92}
\end{equation}
\noindent The brackets $\{\}$ are analogous to brackets in Eq.(62), where the functions $A, B, C$ and $D$ depend on
frequency: $A(\mu)\to A(\mu,\nu)$ etc. According to Eqs.(90) and (91), the functions $A(\mu,\nu), B(\mu,\nu), C(\mu,\nu)$
and $D(\mu,\nu)$ obey the equations (63), where in denominators instead of $(\mu+\mu')$ is to be
taken as $[\alpha(\nu')\mu+\alpha(\nu)\mu']$. Besides, the additional integration over $\nu'$ with the weight function
$\varphi^2(\nu')$ is to be taken. For example, instead of Eq.(66) the frequency dependent analog has the form:

\begin{equation}
2A(\mu,\nu)+B(\mu,\nu)=1+\frac{3}{2}(1-q)\mu\int\limits_0^1d\mu'\int\limits_{-\infty}^{\infty}d\nu'
\frac{\varphi^2(\nu')}{\alpha(\nu')\mu+\alpha(\nu)\mu'}\,T(\mu,\nu;\mu',\nu'),
\label{93}
\end{equation}
\noindent where the function $T(\mu,\nu;\mu',\nu')$ generalize that in Eq.(64). It is of interest to note that
the existence of denominator $[\alpha(\nu')\mu+\alpha(\nu)\mu']$ does not allow to obtain the expression (67)
between $A(\mu,\nu)$ and $B(\mu,\nu)$. Nevertheless, the relation (71) takes place, where

\begin{equation}
a_n(\mu,\nu)=\int\limits_0^1d\mu'\int\limits_{-\infty}^{\infty}d\nu'\frac{\varphi^2(\nu')\mu'^n}
{\alpha(\nu')\mu+\alpha(\nu)\mu'}\,A(\mu',\nu').
\label{94}
\end{equation}
\noindent Analogous expression is valid for function $b_n(\mu,\nu)$. Thus, the relation (72) continues to hold
good for the present case, but the relation (70) does not exist. Contrary to the case of scattering of continuum
radiation, the functions $A,B,C$ and $D$ for the case of scattering of resonant radiation cannot be expressed in
terms of the $H$-functions.

\subsection{The Milne problem for resonant radiation}

The analog of Eq.(22) for resonant radiation is:

\begin{equation}
(\alpha(\nu)-k\mu){\bf g}(\mu,\nu)=\frac{1-q}{2}\int\limits_{-1}^1d\mu'\int\limits_{-\infty}^{\infty}d\nu' \,
\hat P(\mu^2,\mu'^2)\varphi(\nu)\varphi(\nu'){\bf g}(\mu',\nu').
\label{95}
\end{equation}

It follows from this equation that

\begin{equation}
{\bf g}(\mu,\nu)=\frac{\varphi(\nu)}{\alpha(\nu)-k\mu}({\bf g}_0+\mu^2\,{\bf g}_2),
\label{96}
\end{equation}
\noindent where ${\bf g}_0=(a,b)$ and ${\bf g}_2=(c,0)$ are independent of $\mu$ and $\nu$. The equation for
${\bf g}_0+\mu^2{\bf g}_2$ acquires the form:

\begin{equation}
{\bf g}_0+\mu^2{\bf g}_2=(1-q)\int\limits_0^1 d\mu'\Phi(\mu',k)\hat P(\mu^2,\mu'^2)({\bf g}_0+\mu'^2{\bf g}_2).
\label{97}
\end{equation}
\noindent Here
\begin{equation}
\Phi(\mu',k)= \int\limits_{-\infty}^{\infty}d\nu'\frac{\alpha(\nu')\varphi^2(\nu')}{\alpha^2(\nu')-k^2\mu'^2}.
\label{98}
\end{equation}
\noindent Introducing the quantities

\begin{equation}
\Phi_n(k)=\int\limits_0^1d\mu\,\mu^n\,\Phi(\mu,k),
\label{99}
\end{equation}
\noindent we can derive the characteristic equation in the form (83), where instead of $f_n(k)$ one has
to substitute the functions $\Phi_n(k)$. The system of algebraic equations for functions $a, b$ and $c$ coincides with the
system (81) but with the same substitution $f_n(k)\to \Phi_n(k)$.

General formula (27) in the case of resonant line acquires the form:

\[
{\bf I}(0,\mu,\nu)=\frac{Const\, \varphi(\nu)}{\alpha(\nu)-k\mu} \left [\,{\bf g}_0+\mu^2{\bf g}_2-\right.
\]
\begin{equation}
\left.\frac{1}{2\mu}\int\limits_0^1d\mu'\,\int\limits_{-\infty}^{\infty}d\nu'\frac{[\alpha(\nu)-k\mu ]}
{\alpha(\nu')+k\mu'}\,
\hat S_1(\mu,\nu;\mu'\nu')\,\varphi^2(\nu')({\bf g}_0+\mu'^2{\bf g}_2)\right].
\label{100}
\end{equation}

Using the equality

\begin{equation}
\frac{\mu'[\alpha(\nu)-k\mu]}{[\alpha(\nu')+k\mu'][\alpha(\nu')\mu+\alpha(\nu)\mu']}=\frac{1}{\alpha(\nu')+k\mu'}-
\frac{\mu}{\alpha(\nu')\mu+\alpha(\nu)\mu'},
\label{101}
\end{equation}
\noindent Eq.(100) can be written in another form:

\[
{\bf I}(0,\mu,\nu)=\frac{Const\, \varphi(\nu)}{\alpha(\nu)-k\mu}\left [\,{\bf g}_0+\mu^2{\bf g}_2 +
\frac{1}{2}\int\limits_0^1\frac{d\mu'}{\mu'}\int\limits_{-\infty}^{\infty}d\nu'\varphi^2(\nu')
\hat S_1(\mu,\nu;\mu',\nu')({\bf g}_0+\mu'^2{\bf g}_2)-\right.
\]
\begin{equation}
\left.\frac{1}{2\mu}\int\limits_0^1\frac{d\mu'}{\mu'}\int\limits_{-\infty}^{\infty}d\nu'\,
\frac{\varphi^2(\nu')([\alpha(\nu')\mu+\alpha(\nu)\mu']}{\alpha(\nu')+k\mu'}\,\hat S_1(\mu,\nu;\mu',\nu')
({\bf g}_0+\mu'^2{\bf g}_2)\right ].\,\,\,
\label{102}
\end{equation}

\section{Conclusion}

Anisotropy of small grains and molecules gives rise to depolarization of light upon both single and
multiple scattering.
The existence of true absorption of light also changes essentially
the angular distribution and polarization of radiation emerging from an atmosphere. In this paper we consider
the multiple scattering of radiation on freely (chaotic) oriented small particles.

We derived the explicit formulas for intensity and linear polarization of light, reflected from semi-infinite
plane-parallel atmosphere. The standard Milne's problem is also considered. We considered radiative
transfer in both continuum  and  resonant lines. For both types of radiation we used the technique of
invariance principles, which lead to the system of non-linear equation for four H -functions.
We investigated the axially symmetric part of radiation. Only this part depends on depolarization
parameter $\overline{b}_2$.

It is shown that depolarization parameter does not increase the degree of polarization as compared with the
case of pure dipole scattering.
In the case of continuum all four $H$ - functions are expressed in terms of one function. In the case of resonant
radiation only two $H$ - functions, describing the polarization of light, can be expressed  in terms of one function.
For Milne's problems we derived the characteristic equations to calculate unknown  parameter $k$
($I\sim \exp{(k\tau)}$).
These equations depend on usual parameter $\overline{b}_1$, describing the dipole scattering, and on the depolarization
parameter. These equations contain terms proportional to $\overline{b}_1$, $\overline{b}_1^2$, $\overline{b}_2$ and
$\overline{b}_1\overline{b}_2$.

Resonant radiation has the additional effective absorption due to transitions of frequencies from the initial
value to other frequencies. It means that the parameter $k$ is not zero even for conservative atmosphere,
and the outgoing resonant radiation is more elongated  than that for continuum. As a result, the polarisation
of resonant radiation is greater than that in the case of continuum radiation.

The paper is devoted to theoretical investigation of light depolarization due to anisotropy of grains ,
and also due to dipole transitions between molecular levels at resonant scattering.
Recall that
most important effects occur for radiative transfer of resonant radiation. The existence of absorption is
also taken into consideration.

\begin{table}
\caption[]{\small The roots of characteristic equations (23), (24) and (26).}
\label{Table 1}
\setlength{\tabcolsep}{0.2cm}
\centering
\medskip
\begin{tabular}{|l|lllllllll|}
\hline
$\,\,q\,$    &    $0$  & $0.001$ & $0.002$ & $0.003$ & $0.004$ & $0.005$ & $0.010$ & $0.015$ & $0.020$   \\
\hline
$Eq.(23)$   & 0 & 0.054757 & 0.077398 & 0.094754 & 0.109369 & 0.122229 & 0.172511 & 0.210856 & 0.242983   \\

$Eq.(24)$   & 0 & 0.054748 & 0.077391 & 0.094742 & 0.109350 & 0.122202 & 0.172435 & 0.210716 & 0.242768   \\

$Eq.(26)$   & 0 & 0.054743 & 0.077377 & 0.094717 & 0.109311 & 0.122148 & 0.172284 & 0.210445 & 0.242356   \\
\hline
\end{tabular}
\end{table}

\begin{table}
\caption[]{\small The roots of characteristic equations (23), (24) and (26) (continue).}
\label{Table 2}
\setlength{\tabcolsep}{0.2cm}
\centering
\medskip
\begin{tabular}{|l|lllllllll|}
\hline
$q$  & $0.03 $  & $0.04$ & $0.05 $ & $0.1 $ & $0.2$ & $0.3$ & $0.4$ & $0.5$ & $0.6$   \\
\hline
$Eq.(23)$   & 0.296381 & 0.340829 & 0.379485 & 0.525429 & 0.710412 & 0.828635 & 0.907332 & 0.957504 & 0.985624   \\

$Eq.(24)$   & 0.295991 & 0.340233 & 0.378659 & 0.523200 & 0.704828 & 0.819984 & 0.896901 & 0.947380 & 0.978166   \\

$Eq.(26)$   & 0.295255 & 0.339133 & 0.377166 & 0.519583 & 0.697604 & 0.811199 & 0.888707 & 0.941298 & 0.974750   \\
\hline
\end{tabular}
\end{table}

This research was supported by the Program of Prezidium of RAS No21,
the Program of the Department of Physical Sciences of RAS No17, the Federal
Target Program ''Science and Scientific-Pedagogical Personnel of Innovative Russia'' XXXVII in turn
- the action 1.2.1,
and by the Grant from President of the Russian Federation ''The Basic Scientific Schools'' (NSh-1625.2012.2).

The authors are very grateful to an anonymous referee for many useful remarks.


\begin{thebibliography}{20}
\bibitem{1} Abhyankar, K. D., Fymat, A. L. Astrophys. J. Suppl. Ser. {\bf 23}, 35 (1971)
\bibitem{2} Chandrasekhar, S.: Radiative transfer. Dover, New York (1960)
\bibitem{3} Dementyev, A. V. Astr. Lett. {\bf 34}, 574 (2008)
\bibitem{4} Dolginov, A. Z., Gnedin, Yu. N., Silant'ev, N. A. : Propagation and Polarization of
            Radiation in Cosmic Media. Gordon \& Breach Publ., Amsterdam (1995)
\bibitem{5} Faurobert-Scholl, M., Frisch, H. Astron. Astrophys. {\bf 219}, 338 (1989)
\bibitem{6} Horak, H. G., Chandrasekhar, S. Astropys. J. {\bf 134}, 45 (1961)
\bibitem{7} Hummer, D. G. Mon. Not. Roy. Astr. Soc. {\bf 125}, 21 (1962)
\bibitem{8} Ivanov, V. V., Grachev, S. I., Loskutov, V. M. Astron. Astrophys. {\bf 318}, 315 (1997a)
\bibitem{9} Ivanov, V. V., Grachev, S. I., Loskutov, V. M. Astron. Astrophys. {\bf 321}, 968 (1997b)
\bibitem{10} Landi Degl'Innocenti, E., Landolfi, M.: Polarization in spectral lines. Kluwer Acad. Publ.,Dordrecht (2004)
\bibitem{11} Lekht, E. E., Silant'ev, N. A., Rudnitskij G. M., Alexeeva, G. A. Astron. Astrophys. {\bf 492}, 475 (2008)
\bibitem{12} Lenoble, J., J. Quant. Spectrosc. Radiat. Transf. {\bf 10}, 533 (1970)
\bibitem{13} McKenna, S., J. Astrophys. Space Sci. {\bf 108}, 31 (1985)
\bibitem{14} Nagendra, K. N. in ASP Conf. Ser. {\bf 288}, Stellar Atmosphere Modelling, eds. I.Hubeny, D. Mihalas, \&
             K. Werner (San Francisco: ASP), 583 (2003)
\bibitem{15} Nagendra K. N.,\& Sampoorna M. in ASP Conf. Ser. {\bf 405}, Solar Polarization 5, eds. S. V.Berdyugina,
             K. N. Nagendra, \& Ramelli (San Francisco: ASP), 261 (2009)
\bibitem{16} Varshalovich, D. A., Ivanchik, A. V., Babkovskaya N. S. Astr. Lett. {\bf 32}, 29 (2006)
\end{thebibliography}
\end{document}